\documentstyle[aps,preprint,epsfig]{revtex}
\begin{document}
\null{}
\vskip-1.5cm
\hskip12cm{\large HZPP-9802}
\vskip-0.2cm
\hskip12cm{\large Jan 25, 1998}
\vskip1cm
\centerline{The Nonlinear Spatial Damping rate in QGP
\footnote{Work supported by the National Natural Science Fund of China.}}
\centerline{Chen Jisheng\footnote{Permanent address:University of Hydraulic 
\& Electric Engineering/Yichang}\ \ \ Li Jiarong}
\centerline{\small
        \ Institute of particle physics, Central China Normal University}
\centerline{\ Wuhan, 430079, Hubei,P.R. China}

\begin{abstract}
The derivative expansion method has been used to solve the semiclassical kinetic
equations of quark-gluon plasma. The nonlinear spatial damping rate, the imaginary
part of the wave vector, for the longitudinal secondary color waves in the long wavelength limit
 has been calculated numerically.
\end{abstract}
\vspace{0.8cm}
\par
\vspace{0.5cm}
The Landau damping as collisionless damping is an important collective effect in quark-gluon plasma(QGP); 
it describes how the color field is affected by media during its traveling 
in QGP. In the frame of kinetic theory, it's shown that there is no Landau 
damping in linear or Abelian dominance approximation$^{\cite{s1,s2,s3}}$
but there is nonlinear Landau damping$^{\cite{s4}}$. However,
 all the previous works give only the temporal damping rate, i.e., 
the instabilities with complex frequency and real wave 
vector values. As in the electromagnetic plasma$^{\cite{s5}}$, 
if the imaginary part of wave vector for the secondary waves
resulted from the nonlinear interaction in plasmas may not be zero, 
there may be spatial damping also in QGP. The collisionless damping in QGP 
can be completely recognized only when the temporal and spatial 
damping are known. Up to now, neither the QGP kinetic theory 
 nor the finite temperature QCD has given the spatial damping 
 rate directly. In this letter, we will work from the QGP kinetic equations 
 and derive the nonlinear spatial damping rate in QGP finally. 
 \par
 The collisionless kinetic equations for QGP are$^{\cite{s6}}$
\begin{eqnarray}
p^{\mu}D_{\mu}f(x,p)+\frac{1}{2}
p^{\mu}\frac{\partial}{\partial p_{\nu}}\{F_{\mu \nu}
(x),f(x,p)\}=0; \nonumber\\
p^{\mu}D_{\mu}\bar{f}(x,p)-
\frac{1}{2}p^{\mu}\frac{\partial}
{\partial p_{\nu}}\{F_{\mu \nu}(x),\bar{f}(x,p)\}=
0;\\
p^{\mu}\tilde{D}_{\mu}G(x,p)+\frac{1}{2}p^{\mu}\frac{\partial}
{\partial p_{\nu}}\{{\cal F}_{\mu \nu}(x),G(x,p)\}=0\nonumber
\end{eqnarray}
where $f$, $\bar{f}$, $G$ are the distribution functions of quarks, 
antiquarks, gluons in QGP, respectively. $F_{\mu \nu}$, $\cal{F}_{\mu \nu}$ 
 represent the mean field stress tensors in fundamental and 
  adjoint representation, i.e., $F_{\mu \nu}=F_{\mu \nu}^{a}I_a$, 
  ${\cal F}_{\mu \nu} =F_{\mu \nu }^aF_a$. $I_a$ and $F_a$ are 
  the corresponding generators.
  \par
  The mean color field equation coupled with the kinetic equations is 
\begin{eqnarray}
D_{\mu}F^{\mu \nu}(x)=j^{\nu}(x)
\end{eqnarray}
\par
$j^{\nu}$ is the color current
\begin{eqnarray}
j^{\nu}(x)=-\frac{g}{2}\int \frac{d^3p }{(2\pi)^3E_p}p^\nu\left [\left (f(x,p)-\bar{f}(x,p)\right )+2iI_af_{abc}G_{bc}(x,p)\right ]
\end{eqnarray}
where $f_{abc}$ is the structure constant of $SU(N_c)$.
\par
To calculate the imaginary part of the wave vector for the color 
field conveniently, we will solve the QGP kinetic equations 
iteratively in momentum space.
  Noting that the energy density of the color field in QGP is much 
  smaller than the hot energy density in high temperature and 
  density condition, the mean field strength may be selected as the 
  separation of these scales. 
  The derivative expansion method is an effective tool to solve the 
   nonlinear equation$^{\cite{s7,s8}}$. In this method, it's essential that not only 
   the function but also the relevant derivative in equation will be expanded 
   iteratively. 
   To discuss the spatial damping, we will only expand the space derivative 
   in momentum space; the wave vector and the relevant functions
    can be expanded as
\begin{eqnarray}
{\bf k}&=&\sum _{j=0}^{N}\alpha ^j{\bf k}^{(j)};
\nonumber\\
A_i&=&\sum _{j=1}^N \alpha ^j A_i^{(j)}({\bf k}^{(0)},{\bf k}^{(1)},\cdots ,{\bf
k}^{(N)});\\
f&=&\sum _{j=0} ^N \alpha ^j f^{(j)}({\bf k}^{(0)},{\bf k}^{(1)},\cdots ,{\bf
k}^{(N)})\nonumber
\end{eqnarray} 
where $\alpha$ is a dimensionless parameter introduced to denote 
the order of a small quantity. $\bar{f}$ and $G$ can be expanded similarly.
For convenience we work in the temporal gauge, $A_a^0=0$. The relation between the color electrical 
field ${\bf E}$ and the color vector potential ${\bf A}$ can be expressed as:
$E^{i}_a=-\partial A^i_a(x)/\partial t$.
\par
The color current at any order is expanded as
\begin{eqnarray}
j^{h(n)}=-\frac{g}{2}\int \frac{d^3p}{E_p(2\pi)^3}p^{h}\left \{\left (f^{(n)}-\bar{f}^{(n)}\right )+2iI_af_{abc}G_{bc}^{(n)}\right \}
\end{eqnarray} 
\par
From the kinetic equations we can see that the distribution functions will fluctuate
when the mean field is applied to a plasma system as an external field, and the fluctuating
functions reversely influence the mean field. This self-consistent relation between the mean field
and the distribution functions of the plasma particles are well expressed by the 
expansions Eqs.(4) and Eq.(5). 
The leading terms 
describe the linear approximation. The important
dispersion relation obtained in the linear approximation 
will be changed by the nonlinear
interactions of the eigenwaves, i.e., wave vector corrections ${\bf k}^{(1)}$, ${\bf k}^{(2)}$, 
, $\cdots $, ${\bf k}^{(N)}$ will be given to the eigenwave vector ${\bf k}^{(0)}$ correspondingly. As in electromagnetic plasma, if the imaginary parts of the wave vector correction for eigenmodes are not vanishing,
 there will be spatial damping. 
 \par
 To simplify the calculation, we suppose that there's only longitudinal vector
 potential: 
\begin{eqnarray}
A^{i}(k)=k^{(0)i}A(k)/K^{(0)}
\end{eqnarray}
where $K^{(0)}=|{\bf k}^{(0)}|$.
\par
Inserting the expansions Eqs.(4) and Eqs.(5) into the equations (1-2) in momentum space, then equating the coefficients of equal power in $\alpha$ in two sides of these equations, one can obtain a hierarchy of equations.
\par
The first order mean field equation is 
\begin{eqnarray}
-\omega^{2}A^{(1)h}(k)=j^{(1)h}(k)
\end{eqnarray}
The first order transport equation for the distribution function of quarks is
\begin{eqnarray}
p\cdot k^{(0)}f^{(1)}(k,p)&+&\frac{g}{2}\sum_{k_1+k_2=k}(p\cdot k_1^{(0)})\{A_i^{(1)}(k_1),\partial _p^{i}f^{(0)}(k_2,p)\}\nonumber\\
&+&\frac{g}{2}\sum_{k_1+k_2=k}p_i\{A_i^{(1)}(k_1),k_{1\nu}^{(0)}\cdot \partial _p^{\nu}f^{(0)}(k_2,p)\}=0
\end{eqnarray}
where we have defined 
$k^{(0)}$=$(\omega,{\bf k}^{(0)})$ and
\begin{eqnarray}
\sum_{k_1+k_2=k}=\int \frac {d^4k_1d^4k_2}{(2\pi )^4}\delta ^4(k-k_1-k_2). 
\end{eqnarray} 
The first order kinetic equations for antiquarks and gluons are similar to
Eq.(8)
 except the opposite signs of the terms related to $\{ \cdots ,\cdots \}$ for 
 antiquarks, $f$ and $A$ are replaced by $G$ and $\cal A$ for gluons, respectively.
 \par
Assuming background configuration is local neutral, the zero order colorless 
distribution functions can be chosen as the Fermi-Dirac and Bose-Einstein 
equilibrium distribution functions, respectively:
\begin{eqnarray}
f^{(0)}(k,p)=\bar{f}^{(0)}(k,p)=(e^{\beta p\cdot U}+1)^{-1};&~~~~~~~~&
G^{(0)}(k,p)=(e^{\beta p\cdot U}-1)^{-1}
\end{eqnarray}
where $\beta =1/T$ is the inverse of the temperature and $U$ is the local flow velocity(normalized to $U_{\mu}U^{\mu}=1$).
Using Eq.(8) and Eq.(6), we can express the first order distribution functions in terms of the corresponding zero order distribution functions and the first
order field potential
\begin{eqnarray}
f^{(1)}(k,p)=-g\omega\frac{p^{0}k^{(0)i}}{p\cdot k^{(0)}+ip^{0}0^{+}}\frac {df^{(0)}(p)}{dp^{i}}\frac
{A^{(1)}(k)}{K^{(0)}};\nonumber\\
\bar{f}^{(1)}(k,p)=g\omega \frac{p^{0}k^{(0)i}}{p\cdot k^{(0)}+ip^{0}0^{+}}\frac {d\bar{f}^{(0)}(p)}{dp^{i}}\frac
{A^{(1)}(k)}{K^{(0)}};\\
G^{(1)}(k,p)=-g\omega \frac{p^{0}k^{(0)i}}{p\cdot k^{(0)}+ip^{0}0^{+}}\frac {dG^{(0)}(p)}{dp^{i}}\frac {{\cal A}^{(1)}(k)}{K^{(0)}}\nonumber
\end{eqnarray}
By inserting Eq.(11) into the mean field equation Eq.(7) we can obtain the dispersion relation satisfied by the frequency 
 and wave vector of the eigenwave in the first order approximation.
\begin{eqnarray}
\epsilon (\omega,{\bf k}^{(0)})=1+
\frac {3\omega_p^{2}}{K^{(0)2}}
\left [
1-\frac {\omega }{2K^{(0)}}
\left (\ln\left |\frac{K^{(0)}+\omega }{K^{(0)}-\omega}\right |-i\pi \theta (K^{(0)}-\omega )\right )
\right ]=0
\end{eqnarray}
and the solution is 
\begin{eqnarray}
A^{(1)\sigma}_{{\bf k}^{(0)}}=-i\frac{\pi}{\omega }
E^{\sigma}_{{\bf k}^{(0)}}
\left [e^{-i\phi ^{\sigma}_{{\bf k}^{(0)}}}
\delta(\omega-\omega^{\sigma}_{{\bf k}^{(0)}})+
e^{i\phi ^{\sigma}_{{\bf k}^{(0)}}}
\delta(\omega+\omega^{\sigma}_{{\bf k}^{(0)}})\right ]
\end{eqnarray}
where $E^{\sigma}_{{\bf k}^{(0)}}$ and $\phi ^{\sigma}_{{\bf k}^{(0)}}$ 
are the initial amplitude and phase of the oscillation, respectively.
$\omega_p=\sqrt{(2N_c+N_f)g^{2}T^{2}/18}$ is the plasma frequency,
 with $N_f$ being the number of flavors for quarks. The dispersion relation Eq.(12)
agrees with the leading order result using the hard thermal loops in finite temperature QCD
 and the classical nature of the hard thermal loops has
been investigated extensively by Blaizot and Jancu,
etc.$^{\cite{z1,z2}}$. As shown by U.~Heinz,
 the eigenwaves
satisfying Eq.(12) are always timelike, i.e., $\frac {\omega }{K^{(0)}}>1
$(we set $\hbar =c=1$ for convenience). It means that the phase velocity of
eigenwaves is bigger than the velocity of light. So these waves can't
exchange energy with plasma particles and do not undergo damping in the linear
approximation$^{\cite{s1,s2}}$. 
In the long-wavelength region, the dispersion relation Eq.(12) is reduced to the 
form 
\begin{eqnarray}
\omega ^2=\omega _p^2+\frac {3}{5} K^{(0)^{2}}
\end{eqnarray}
\par
The following identities will be used to simplify the color current contributed by gluons in the following discussion: 
\begin{eqnarray}
&~&f_{abc}f_{abd}=N_c\delta _{cd};~~~~~~~~
if_{abc}G_{bc}=tr(F_aG); ~~~~~~~~
[F_a,F_b]=if_{abc}F_c;\nonumber\\
&~&I_dtr(F_d[F_a,F_b])=N_c[I_a,I_b];~~~~~~~~
I^{d}tr(F^{d}[F^{a},[F^{b},F^{c}]])=N_c[I^{a},[I^{b},I^{c}]];\nonumber\\
&~~~~~&\rm tr(\{F_a,\{F_b,F_c\}\}F_d\})=4\delta _{ad} \delta _{cb}+2\delta _{ab}\delta _{cd}
+2\delta _{ac} \delta _{bd}+N_cd^{ade}d^{bce}
\end{eqnarray}
\par
The second order equations can be discussed  in the same manner.
Substituting the results in the expression for the second color current 
by using the distribution functions obtained from the second order kinetic equations, 
one can 
find the second order mean field equation is reduced to 
\begin{eqnarray}
-\omega^{2}\epsilon(\omega,{\bf k}^{(0)})A^{(2)}(k)
&=&\int\frac {d^3p}{(2\pi )^3}\left [N_f\frac{d}{dE_p}
\left (f^{(0)}(p)+\bar{f}^{(0)}(p)\right )+2N_c \frac {dG^{(0)}(p)}{dE_p}\right ]\nonumber\\
&~&\left \{g^3 \frac{1}{p\cdot k^{(0)}+ip^{0}0^{+}}\frac {{\bf p}\cdot {\bf k^{(0)}}}
{K^{(0)}}\sum_{k_1+k_2=k}\frac {\omega _2}{p\cdot k_2+ip^{0}0^{+}}
\frac{{\bf p}\cdot {\bf k_1}^{(0)}}{K_1^{(0)}}\frac{{\bf p}\cdot {\bf k_2}^{(0)}}{K_2^{(0)}}\right.\nonumber\\
&~&\left.[A^{(1)}(k_1),A^{(1)}(k_2)]
-g^2 \frac{{\bf p} \cdot {\bf k}^{(1)}}{p^0}
\frac{A^{(1)}(k)}{(p\cdot k^{(0)}+ip^0 0^+)^2}
\frac{({\bf p}\cdot{\bf k}^{(0)})^3}{(K^{(0)})^2}\right \}
\end{eqnarray}
\par
Eq.(16) describes the three-wave processes owing to the nonlinear
coupling term $A^{(1)}(k_1)A^{(1)}(k_2)$ of 
two eigenwaves with wave vectors ${\bf k}_1$ and ${\bf k}_2$ 
into a secondary wave with
wave vector ${\bf k}$.
However, as pointed out in the Refs.$\cite{s9,s10}$, the three-wave processes 
are forbidden,
there will be nonlinear effects of the three-wave processes on wave vectors
 only in higher order perturbation.
\par
Now let's discuss the third order equations. 
The third order field equation is 
\begin{eqnarray}
-\omega^{2}A^{(3)}(k)
\frac{k^{(0)h}}{K^{(0)}}&+&g\sum_{k_1+k_2=k}\frac{k^{(1)i}k_1^{(0)i}k_2^{(0)h}}{K_1^{(0)}K_2^{(0)}}
[A^{(1)}(k_1),A^{(1)}(k_2)]\nonumber\\
&+&g^{2}\sum_{k_1+k_2=k}\sum_{k_3+k_4=k_2}\frac{k^{(0)i}k_3^{(0)i}k_4^{(0)h}}
{K_1^{(0)}K_2^{(0)}K_4^{(0)}}[A^{(1)}(k_1),[A^{(1)}(k_3),A^{(1)}(k_3)]]=j^{(3)h}(k)
\end{eqnarray}
and the third order kinetic equation for quark is 
\begin{eqnarray}
p\cdot&& k^{(0)}f^{(3)}(k,p)-k_i^{(2)}p_if^{(1)}(k,p)-gp_i\sum_{k_1+k_2=k}[A_i^{(1)}(k_1),f^{(2)}(k_2,p)]-p_ik_i^{(1)}f^{(2)}(k,p)\nonumber\\
&~&-gp_i\sum_{k_1+k_2=k}[A_i^{(2)}(k_1),f^{(1)}(k_2,p)]+\frac{g}{2}\sum_{k_1+k_2=k}p\cdot k_1^{(0)}\{A_i^{(1)}(k_1),\partial _p^{i}f^{(2)}(k_2,p)\}\nonumber\\
&~&+\frac{g}{2}\sum_{k_1+k_2=k}p\cdot k_1^{(0)}\{A_i^{(2)}(k_1),\partial _{p}^{i}f^{(1)}(k_2,p)\}-\frac{g}{2}\sum_{k_1+k_2=k}p_ik_{1i}^{(1)}\{A_j^{(2)}(k_1),\partial _{p}^{j}f^{(0)}(k_2,p)\}\nonumber\\
&~&+\frac{g}{2}\sum_{k_1+k_2=k}p_i\{A_i^{(1)}(k_1),k_{1\nu}^{(0)}\partial _{p}^{\nu }f^{(2)}(k_2,p)\}+\frac{g}{2}\sum_{k_1+k_2=k}p_i\{A_i^{(2)}(k_1),k_{1\nu}^{(0)}\partial _{p}^{\nu }f^{(1)}(k_2,p)\}\nonumber\\
&~&+\frac {g}{2}\sum_{k_1+k_2=k}p_i\{A_i^{(2)},k_{1\nu}^{(1)}\partial _p^{\nu }f^{(0)}(k_2,p)\}+\frac {g}{2}\sum_{k_1+k_2=k}(p\cdot k)\{A_i^{(3)}(k_1),\partial _p^if^{(0)}(k_2,p)\}\nonumber\\
&~&-\frac {g}{2}\sum_{k_1+k_2=k}p_ik_{1i}^{(1)}\{A_j^{(2)}(k_1),\partial _p^jf^{(0)}(k_2,p)\}-\frac {g}{2}\sum_{k_1+k_2=k}p_ik_{1i}^{(2)}\{A_j^{(1)}(k_1),\partial _p^jf^{(0)}(k_2,p)\}\nonumber\\
&~&+\frac {g}{2}\sum_{k_1+k_2=k}p_i\{A_i^{(1)}(k_1),k_{1\nu}^{(1)}\partial _p^{\nu }f^{(1)}(k_2,p)\}+\frac {g}{2}\sum_{k_1+k_2=k}p_i\{A_i^{(1)}(k_1),k_{1\nu }^{(2)}\partial _p^{\nu }f^{(0)}(k_2,p)\}\nonumber\\
&~&+\frac{g}{2}\sum_{k_1+k_2=k}p_i\{A_i^{(3)}(k_1),k_{1 \nu }^{(0)} \partial _{p}^{\nu }f^{(0)}(k_2,p)\}\nonumber\\
&~&-\frac {g^2}{2}\sum_{k_1+k_2=k}\sum_{k_3+k_4=k_1}p_i\{[A_i^{(1)}(k_3),A_j^{(2)}(k_4)],\partial _p^jf^{(0)}(k_2,p)\}\nonumber\\
&~&-\frac{g^2}{2}\sum_{k_1+k_2=k}\sum_{k_3+k_4=k_1}p_i\{[A_i^{(1)}(k_3),A_j^{(1)}(k_4)],\partial _{p}^{j}f^{(1)}(k_2,p)\}\nonumber\\
&~&-\frac {g^2}{2}
\sum_{k_1+k_2=k}\sum_{k_3+k_4=k_1}p_i\{[A_i^{(2)}(k_3),A_j^{(1)}(k_4)],\partial _p^jf^{(0)}(k_2,p)\}=0
\end{eqnarray}
The third order kinetic equations for anti-quarks and gluons are similar 
except for the changes pointed out in Eq.(8).
\par
Analogously to the electromagnetic plasma theory$^{\cite{s11}}$, the wave vector ${\bf k}$ can be expressed as ${\bf k}=\vec{\beta}-i\vec{\alpha }$, where
$\beta=|\vec{\beta}|$ is the phase constant and $\alpha=|\vec{\alpha }|$ is damping  constant, i.e., the amplitude of the eigenwave will exponentially 
decrease in the direction of 
$\vec{\alpha}$. The nonlinear spatial damping rate at this order is defined as 
\begin{eqnarray}
\alpha=-Im|{\bf k}^{(2)}|=-ImK^{(2)}.
\end{eqnarray}
\par
As the oscillations are developed from random thermal motions, we have $\langle A^{(1)a}(k)\rangle=0$, where $\langle~~\rangle$
means the average with respect to the random phase of the oscillations. So we
obtain
\begin{eqnarray}
\langle A^{(1)a}(k)A^{(1)b*}(k')\rangle=(2\pi )^4
\delta ^4(k-k')\delta _{a}^{b}\langle A^{(1)2}(k)\rangle_{{\bf k}^{(0)}\omega
};\nonumber\\
\langle A^{(1)2}(k)\rangle_{{\bf k}^{(0)}\omega }=\frac                                                  
{\pi}{\omega ^{2}}[\delta (\omega-\omega_{{\bf k}^{(0)}})]I_{{\bf k}^{(0)}};~~~~~I_{{\bf k}^{(0)}}=\frac {|E_{{\bf k}^{(0)}}|}{2V} 
\end{eqnarray}
where $V$ is the volume of the plasma; $I_{{\bf k}^{(0)}}$ characterizes the total intensity of the fluctuating oscillation with 
frequency $\omega_{{\bf k}}$ and $-\omega_{{\bf k}}$. In an equilibrium plasma and the long wavelength limit($K^{(0)}=0$) 
there are $I_{{\bf k}^{(0)}}=4\pi T$. 
The average of the product of the field potential can be expanded
as$^{\cite{s7,s12}}$
\begin{eqnarray}
\langle A^{(1)}(k_1)A^{(1)}(k_2)A^{(1)}(k_3)A^{(1)}(k_{4 })\rangle&=&
\langle A^{(1)}(k_1)A^{(1)}(k_2)\rangle\langle A^{(1)}(k_3)A^{(1)}(k_4)\rangle\nonumber\\
&~&+\langle A^{(1)}(k_1)A^{(1)}(k_3)\rangle\langle A^{(1)}(k_2)A^{(1)}(k_4)\rangle\nonumber\\
&~&+\langle A^{(1)}(k_1)A^{(1)}(k_4)\rangle\langle A^{(1)}(k_2)A^{(1)}(k_3)\rangle
\end{eqnarray}
\par
By inserting the third order distribution functions Eqs.(18) into Eq.(17), multiplying both sides of the equation by $A^{(1)b}(k')$
 and performing the average of the results with respect to the random phase, 
 we have 
\begin{eqnarray}
-Im\int \frac {d^3p}{E_p(2\pi )^3}\left (\frac {{\bf p}\cdot {\bf k}^{(0)}}{K^{(0)}}\right )^2
\frac {{\bf p}\cdot {\bf k^{(2)}}}{\left ( p\cdot k^{(0)}+ip^00^+ \right )^2}\left [N_f\frac {df^{(0)}(p)}{dE_P}+N_c\frac {dG^{(0)}(p)}{dE_p}\right ]
=A_1+A_2;
\end{eqnarray}
\begin{eqnarray}
A_1=&Im&\left (-\frac {g^2}{\omega _p}\int \frac {d^3p}{(2\pi )^3E_p}
\int \frac {d^4k_1}{(2\pi )^4}\langle A^{(0)^2}\rangle _{{\bf k}_1^{(0)}\omega}
\left (\frac {\omega _1}{K_1^{(0)}}\right )^2
\left (\frac {\omega}{K^{(0)}} \right)^2 \right.\nonumber\\
&\times &\left.\frac {1}{p\cdot (k^{(0)}-k^{(0)}_1)+ip^00^+}
\left (\frac {p^0{\bf k}^{(0)}\cdot {\bf k}^{(0)}_1}
{p\cdot k^{(0)}+ip^00^+}-{\bf p}\cdot {\bf k}^{(0)}
\frac {\omega {\bf p}\cdot {\bf k}^{(0)}_1-p^0{\bf k}^{(0)}\cdot {\bf k}^{(0)}_1}
{(p\cdot k^{(0)}+ip^00^+)^2} \right )\right. \nonumber\\ 
&\times& \left.\left\{\left [{\bf k}^{(0)}\cdot {\bf k_1^{(0)}}
\left (
       \frac {1}{p\cdot k_1^{(0)}+ip^00^+}
       -\frac {1}{p\cdot k^{(0)}+ip^00^+}
\right )
       +({\bf p}\cdot {\bf k_1}^{(0)})
\frac {-\omega _1{\bf p}\cdot {\bf k}^{(0)}/p^0+{\bf k}^{(0)}\cdot {\bf k}_1^{(0)}}
{({\bf p}\cdot {\bf k}_1^{(0)}+ip^00^+)^2}\right. \right.\right. \nonumber\\
&-&\left.\left.\left.({\bf p}\cdot {\bf k})
\frac {-\omega {\bf p}\cdot {\bf k}_1^{(0)}/p^0+{\bf k}^{(0)}\cdot {\bf k}_1^{(0)}}
{(p\cdot k^{(0)}+ip^00^+)^2}\right ]
\left (\frac {7}{6}N_f\frac {df^{(0)}(p)}{dE_p}+\frac {22}{3}\frac {dG^{(0)}(p)}{dE_p} \right )
\right.\right.\nonumber\\
&+&\left.\left. ({\bf p}\cdot {\bf k}_1^{(0)})\frac {({\bf p}\cdot {\bf k}^{(0)})}{p^0}
\left ( 
        \frac {1}{p\cdot k_1^{(0)}}-\frac {1}{p\cdot k^{(0)}+ip^00^+}
\right )
\left (\frac {7}{6}N_f\frac {d^2f^{(0)}(p)}{dE_p^2}+\frac {22}{3}\frac {d^2G^{(0)}(p)}{dE_p^2} \right )
\right\}\right );\nonumber\\
A_2=&Im& \left \{\int \frac {d^3p}{(2\pi )^3E_p}\int \frac {d^4k_1}{(2\pi )^4}
\langle A^{(1)^{2}} \rangle_{{\bf k}_1^{(0)}\omega}
\left (N_f\frac {df^{(0)}}{dE_p}+N_c\frac {dG^{(0)}(p)}{dE_p}\right )\right.\nonumber\\
&\times &\left. \frac {1}{p\cdot k^{(0)}+ip^00^+}\times 
\left [ \frac {-6g^2}{\omega _p}\left (\frac {{\bf p}\cdot {\bf k}^{(0)}}{K^{(0)}}\right )^2\left (
\frac {{\bf p}\cdot {\bf k}^{(0)}}{K^{(0)}_1}
\right )^2\frac {1}{p\cdot (k^{(0)}-k^{(0)}_1)+ip^00^+}\right.\right.\nonumber\\
&\times & \left.\left.\left (\frac {\omega _1}{p\cdot k_1+ip^00^+}-\frac {\omega}{p\cdot k^{(0)}+ip^00^+}\right )
+\frac {12g^4}{\omega _p}\left (\frac {{\bf p}\cdot {\bf k}^{(0)}}{K^{(0)}}\right )
\left (\frac {{\bf p}\cdot {\bf k}_1^{(0)}}{K_1^{(0)}}\right )
\left (\frac {{\bf p}\cdot ({\bf k}^{(0)}-{\bf k}_1^{(0)})}
{|{\bf k}^{(0)}-{\bf k}_1^{(0)}|} \right )\right.\right.\nonumber\\
&\times &\left.\left.
\left (\frac {\omega _1}{p\cdot k_1+ip^00^+}-\frac {\omega }{p\cdot k^{(0)}+ip^00^+} \right )
\times \int \frac {d^3p'}{(2\pi )^3E_p'}\left (N_f\frac {df^{(0)}(p')}{dE_p'}
+N_c\frac {dG^{(0)}(p')}{dE_p'}\right )\right.\right.\nonumber\\
&\times &\left.\left.
\left (
        \frac {{\bf p}'\cdot ({\bf k}^{(0)}-{\bf k}_1^{(0)})}{|{\bf k}^{(0)}-{\bf k}_1^{(0)}|}
\right )
\left (
        \frac {{\bf p}'\cdot {\bf k}^{(0)}}{K^{(0)}}
\right )
\left (
        \frac {{\bf p}'\cdot {\bf k}_1^{(0)}}{K_1^{(0)}}
\right )
\frac {1}{(\omega -\omega _1)^2}\frac {1}{\epsilon (\omega -\omega _1,{\bf k}^{(0)}-{\bf k}_1^{(0)})}\right.\right.\nonumber\\
&\times&\left.\left.
\frac {1}{p'\cdot (k^{(0)}-k_1^{(0)})+ip^00^+}
\left (
        \frac {\omega _1}{p'\cdot k_1^{(0)}+ip^00^+}-\frac {\omega}{p'\cdot k^{(0)}+ip'^00^+}
\right )
\right ]
\right \}\nonumber
\end{eqnarray}
\par
To get the numerical result of the nonlinear spatial damping rate, we
 perform the integrals in cylindrical coordinates and in the local rest frame of the plasma particle. 
 \par
At first the left hand 
 side of Eq.(22) can be simplified by averaging over all directions of ${\bf k}^{(2)}$ and selecting the direction of 
 ${\bf k}^{(0)}$ as the direction of the polar axis, i.e., we have 
 \begin{eqnarray}
 -Im\int \frac {d^3p}{E_p(2\pi )^3}
 \frac {({\bf p}\cdot {\bf k}^{(0)})^2}{(K^{(0)})^2}
 \frac {{\bf p}\cdot {\bf k}^{(2)}}{(p\cdot k^{(0)}+ip^00^+)^2}
 \left (N_f\frac
 {d}{dE_p}f^{(0)}(p)+N_c\frac {dG^{(0)}(p)}{dE_p}\right )\nonumber\\
=-Im \langle K^{2} \rangle \frac {1}{3\omega_p^2}
\int \frac {d^3p}{E_p(2\pi )^3}\cos^2 \phi 
\left ( N_f\frac {d}{dE_p}f^{(0)}(p)+N_c\frac {dG^{(0)}(p)}{dE_p}\right )
 \end{eqnarray}
where $\phi $ is the angle between ${\bf p}$ and ${\bf k}^{(0)}$. 
$\langle ~~ \rangle $ means the average 
over all directions of ${\bf k}^{(2)}$$^{\cite{s12,s13}}$.
\par
Now we can see that $Im\langle K^{(2)}\rangle$ can be extracted from the left 
hand side of Eq.(22), and the numerical value of $Im\langle K^{(2)}\rangle$
 is determined by $A_1$ and $A_2$. Before performing the integrals in $A_1$
  and $A_2$, we analyze the mechanism of the nonlinear spatial damping.
  The important relation 
$p^{0}/(p\cdot k'+ip^{0}0^{+})=P\left [1/(\omega '-{\bf v}\cdot {\bf k}'\right ]-i\pi \delta (\omega '-{\bf v}\cdot {\bf k}')$
is very useful in our discussion, where ${\bf v}={\bf p}/p^0$ is the
velocity of particle and $P$ stand for the principal
 value of the function. We can see that
 $Im\langle K^{(2)}\rangle$ is not vanishing only when at least one of the
 imaginary parts of $1/(p\cdot k^{(0)}+ip^00^+)$, $1/(p\cdot
 k_1^{(0)}+ip^00^+)$ and $1/(p\cdot (k^{(0)}-k_1^{(0)})+ip^00^+)$ is not
 zero. As is shown in the linear approximation, the color eigenwaves are
 timelike, i.e., the phase velocity of waves cann't approach the velocity of
particle and the imaginary parts of $1/(p\cdot k^{(0)}+ip^00^+)$ and $1/(p\cdot
 k_1^{(0)}+ip^00^+)$
are zero. The term $1/(p\cdot (k^{(0)}-k_1^{(0)})+ip^00^+)$ appearing in $A_1$ 
and $A_2$ are very crucial, it describes that the eigenwaves with wave
vectors ${\bf k}^{(0)}$ and ${\bf k}^{(0)}_1$ may produce the 
secondary waves with wave vectors ${\bf k}^{(0)}-{\bf k}^{(0)}_1$
through the nonlinear interactions. Even if the eigenwaves are   
timelike, these secondary waves may be spacelike, i.e., their phase velocity $\frac
{\omega-\omega _1}{{\bf k}^{(0)}-{\bf k}^{(0)}_1}$ may be smaller than the
velocity of light and can approach the velocity of particle.
So the imaginary part of $\delta \left ((\omega -\omega _1)-{\bf v}\cdot ({\bf k}^{(0)}-{\bf k}^{(0)}_1 \right
)$ may not be zero and the secondary waves may exchange energy with plasma particles and be damped  
 by the particles. It should be emphasized that though the mechanism is
analogous to the mechanism of nonlinear Landau damping in eletromagnetic
plasmas, the nonlinear coupling of the waves takes place through the
nonlinear relations $\left [A^{(1)}(k_1),A^{(1)}(k_2)\right ]$ etc. in
Eq.(17), i.e., it describes the nonlinear and non-Abelian characteristics
of QGP.
 \par
  The similar integral to $A_1$ and $A_2$ has been discussed in Ref.\cite{s4}, by taking 
  only the leading order in $g$ and taking the long-wavelength limit, we find
\begin{eqnarray}  
A_1+A_2\sim 0.42 T.
\end{eqnarray}
\par
After finishing the integral of the right hand side of Eq.(23),
 and from Eq.(24), we obtain the nonlinear
spatial damping rate for a pure gluon gas in the long  wavelength limit
\begin{eqnarray}  
\alpha =-Im\langle K^{(2)} \rangle \sim 2.52 g^2T
\end{eqnarray}
\par
Now we summarize briefly. In this paper, we use the derivative expansion method to study 
the nonlinear and the non-Abelian effects given
by the kinetic equations of QGP. By solving the 
equations to the third order, the numerical result of the spatial 
damping rate  $\alpha$ for the secondary waves in the long wavelength limit 
is obtained. This result indicates that the amplitude of the secondary waves in 
QGP decrease exponentially, and its damping rate is $2.52 g^2T$ Neper approximately.
 It makes us have a concrete understanding about the spatial traveling of
  color waves in QGP. 

\end{document}